\documentclass[12pt]{article}
 \pdfoutput=1
 \usepackage{graphicx}
  \usepackage{hyperref}
  \usepackage{epsfig}
  \usepackage{amsmath}
  \usepackage{amssymb}
  \usepackage{color}
  \usepackage{mciteplus}
  \newlength{\abstractwidth}
  \newcommand{\be}{\begin{equation}}
  \newcommand{\ee}{\end{equation}}

  \renewcommand{\title}[1]{\vbox{\center\bf{\Large{#1}}}\vspace{5mm}}
  \renewcommand{\author}[1]{\vbox{\center#1}\vspace{5mm}}
  \newcommand{\address}[1]{\vbox{\center\em#1}}
  \newcommand{\email}[1]{\vbox{\center\tt#1}\vspace{5mm}}
  \definecolor{darkgreen}{rgb}{0,.5,0}

\begin{document}

\begin{titlepage}
\rightline{MIT-CTP/4626}
\begin{center}
\hfill \\
\hfill \\
\vskip 1cm

\title{Two-dimensional conformal field theory and the butterfly effect}

\author{Daniel A. Roberts${}^a$ and Douglas Stanford${}^{b}$}

\address{$^{a}$ Center for Theoretical Physics {\it and} \\  Department of Physics, Massachusetts Institute of Technology \\
Cambridge, MA, USA  

\vspace{10pt}

$^{b}$ School of Natural Sciences, Institute for Advanced Study,\\ Princeton, NJ, USA}

\email{drob@mit.edu \hspace{20pt} stanford@ias.edu}

\end{center}

  \begin{abstract}
We study chaotic dynamics in two-dimensional conformal field theory through out-of-time order thermal correlators of the form $\langle W(t)VW(t)V\rangle$. We reproduce bulk calculations similar to those of \cite{Shenker:2013pqa}, by studying the large $c$ Virasoro identity block. The contribution of this block to the above correlation function begins to decrease exponentially after a delay of $\sim t_* - \frac{\beta}{2\pi}\log \beta^2E_w E_v$, where $t_*$ is the scrambling time $\frac{\beta}{2\pi}\log c$, and $E_w,E_v$ are the energy scales of the $W,V$ operators.
  \end{abstract}
  \end{titlepage}

\tableofcontents

\baselineskip=17.63pt

\section{Introduction}
Chaos is a generic feature of thermal systems, but it is difficult to study directly. This is both because chaotic systems tend to be poorly suited to perturbation theory, and because e.g. the butterfly effect is not easily visible in time-ordered vacuum correlation functions. In recent work  \cite{Shenker:2013pqa,Shenker:2013yza,Leichenauer:2014nxa,Roberts:2014isa,kitaev}, gauge/gravity duality was used to expose chaos in the gauge theory in terms of shock waves on the horizons of AdS black holes. 

In this paper, we will reproduce part of that analysis without using holography directly. We will instead focus on a particular contribution to the four point function given by the large $c$ Virasoro identity block. The close relationship between 2+1 gravity and the identity Virasoro block has been demonstrated recently in \cite{Hartman:2013mia,Fitzpatrick:2014vua,Asplund:2014coa}; our work should be understood as an application of the techniques in these papers to the problem studied in \cite{Shenker:2013pqa,Shenker:2013yza,Leichenauer:2014nxa,Roberts:2014isa,kitaev}.

The central object of our study will be a non-time-ordered correlator of the form\footnote{In this paper, we use the standard convention $W(t) = e^{iHt}W e^{-iHt}$.}
\be
\langle W(t) V W(t) V\rangle_\beta \label{WVWV}.
\ee
Here, the subscript $\beta$ indicates a thermal expectation value. $W,V$ are approximately local operators, smeared over a thermal scale, and with one-point functions subtracted. In a suitably chaotic system, this type of correlation function becomes small as $t$ becomes large. We believe that this happens for practically any choice of $W$ and $V$ (subject to the above conditions) and that, in fact, this behavior is a basic diagnostic of quantum chaos. To motivate this point, it is helpful to regard Eq.~(\ref{WVWV}) as an inner product of two states: one given by applying $V$ then $W(t)$, and one given by applying $W(t)$ then $V$. Intuition from classical chaos suggests that these states should be rather different if the time $t$ is sufficiently long, hence a small overlap. This was confirmed for holographic theories in the gravity analysis of \cite{Shenker:2013pqa,Shenker:2013yza,Roberts:2014isa}. Such correlation functions have also been studied recently, using similar techniques, by Kitaev \cite{kitaev}, who interpreted the initially exponential decrease in terms of Lyapunov exponents, following \cite{larkin}. Kitaev also pointed out the unusual quantization of these exponents in a theory dual to gravity.

The discussion above should be contrasted with the behavior of correlation functions with the orderings
\be
\langle W(t) W(t) V V\rangle_\beta, \hspace{20pt} \langle V W(t) W(t) V\rangle_\beta,\label{WWVV}
\ee
both of which tend to an order-one value $\langle W W\rangle_\beta\langle V V\rangle_\beta$ as $t$ becomes large. In the case of the $WWVV$ configuration, this is because the correlator is time-ordered, and the connected contribution decays with $t$. The $VWWV$ ordering can be understood as an expectation value of $WW$ in a state given by acting with $V$ on the thermal state. If the energy injected by $V$ is small, the state will relax and the expectation value will approach the thermal value, $\langle WW\rangle_\beta$, multiplied by the norm of the state, $\langle V V\rangle_\beta$.

Each of the Lorentzian correlators in (\ref{WVWV}) and (\ref{WWVV}) can be obtained by analytic continuation of the same Euclidean four point function. The sharp difference in behavior arises because the continuation defines a multivalued function, with different orderings corresponding to different sheets. To see the buttefly effect, one has to move off the principal sheet.

Although what we have said up to now is completely general, most of the paper will be restricted to two dimensional conformal field theory. There are two reasons for this. The first is that conformal symmetry in two dimensions relates thermal expectation values (in a spatially infinite system) to vacuum expectation values. This makes it possible to study the thermal correlation functions described above using standard conformal blocks on the plane. A key feature of the conformal mapping is that long time $t$ in (\ref{WVWV}) and (\ref{WWVV}) translates to small values of the cross ratios $z, \bar{z}$. This suggests an OPE limit, and indeed, for the orderings in Eq.~(\ref{WWVV}), the large $t$ behavior is dominated by the identity operator in the operator product of $WW$. However, for the ordering in Eq.~(\ref{WVWV}), $z$ has moved to a second sheet of the Euclidean correlation function, where the ordinary OPE does not apply.

A very similar situation arose in the study of high energy AdS scattering by \cite{Cornalba:2006xk,Cornalba:2006xm,Cornalba:2007zb}. As pointed out in those references, a formal expansion in continued conformal blocks is still possible on the second sheet of the four point function. The conformal block associated to global primaries with spin $J$ and energy $E$ give a contribution proportional to (in our variables)
\be
e^{\frac{2\pi}{\beta}(J-1)|t|}e^{-\frac{2\pi}{\beta}(E-1)|x|},
\ee
where $x$ is the spatial separation of the operators $W,V$. Here, the contribution of high-spin operators increases with $t$. In purely CFT terms, this is the reason the ordering in Eq.~(\ref{WVWV}) is more interesting than those in Eq.~(\ref{WWVV}). 

This brings us to the second advantage of 2d CFT: we can sum a particular class of global primaries by studying the Virasoro conformal block of the identity operator. As we will see, this reproduces an AdS${}_3$ bulk calculation in the style of \cite{Shenker:2013pqa,Shenker:2013yza,Roberts:2014isa}, and it provides a purely field-theoretic window into fast scrambling \cite{Hayden:2007cs,Sekino:2008he,Lashkari:2011yi,Brown:2012gy}. However, we emphasize that contributions of operators with high spin are important. Even with the usual assumption of a large gap in operator dimensions, restricting to Virasoro descendants of the identity becomes a bad approximation at sufficiently large $t$. In the bulk, what is left out are stringy corrections to the Einstein gravity computation \cite{stringy}. Still, the Einstein gravity results provided a useful starting point for the investigation of chaos using holography, and we hope that the Virasoro identity block will provide a similarly useful starting point in conformal field theory.

The plan of the paper is as folllows. In \S~\ref{S2}, we will set up the configuration of operators and review the Euclidean conformal block expansion. In \S~\ref{S3}, we will discuss the analytic continuations required to obtain Eq.s~(\ref{WVWV}) and (\ref{WWVV}) from the Euclidean correlator. We will also review the ``second sheet OPE,'' following \cite{Cornalba:2006xm}. In \S~\ref{S4}, we will study the contribution of the stress tensor (and its Virasoro module) in detail. By analytically continuing Virasoro conformal blocks, we will reproduce a special case of the gravity calculations that demonstrated the butterfly effect in \cite{Shenker:2013pqa,Shenker:2013yza,Roberts:2014isa}. For the convenience of the reader, this calculation is spelled out in detail in appendix \ref{bulk}.

Before we begin, we will clarify one further point. We have advertised the behavior of the correlation function (\ref{WVWV}) as a diagnostic of quantum chaos. We should contrast this with the expected behavior in an integrable system. In such a system, certain out-of-time order correlation functions of the type (\ref{WVWV}) might tend to zero for large time $t$. However, this should not happen for {\it all} pairs of operators $W,V$. In appendix \ref{ising} we demonstrate this in a familiar integrable system: the two dimensional Ising model.

\paragraph{Note:}
While this work was in progress, we became aware of a very interesting project by S. Jackson, L. McGough, and H. Verlinde \cite{Jackson:2014nla}. We believe that their work is closely related to ours, and we have arranged to post our articles simultaneously.

\section{CFT calculations}

\subsection{Conventions and review} \label{S2}
\begin{figure}
\begin{center}
\includegraphics[scale=.8]{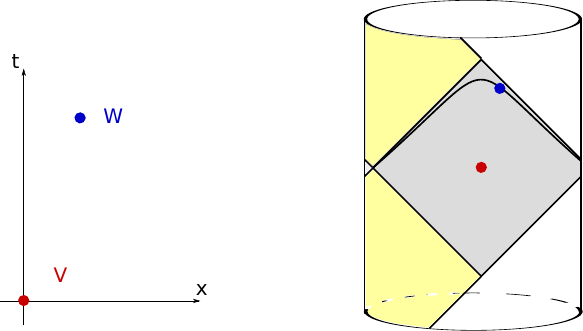}
\caption{{\bf Left:} the spacetime arrangement of the $W$ and $V$ operators. {\bf Right:} their locations after the conformal mapping, viewed in the Rindler patch on the boundary of AdS${}_3$ (grey) covered by $x,t$. The union of the grey and yellow regions are the Poincare patch covered by $z,\bar{z}$.}\label{figtwo}
\end{center}
\end{figure}

In this paper, we will study thermal four-point correlation functions of $W$ and $V$, Eq.s (\ref{WVWV}) and (\ref{WWVV}). Eventually, these operators will be arranged in the timelike configuration shown in Fig.~\ref{figtwo}, where $V$ is at the origin, and $W$ is at position $t >x >0$. However, we will obtain these correlation functions by starting with the Euclidean correlator and analytically continuing. 
We will work exclusively in the setting of two-dimensional conformal field theory. This allows us to map the thermal expectation values to vacuum expectation values through the conformal transformation
\be
z(x,t) = e^{\frac{2\pi}{\beta}(x+t)}, \hspace{20pt} \bar{z}(x,t) = e^{\frac{2\pi}{\beta}(x-t)}.\label{z}
\ee
Here, $x,t$ are the original coordinates on the spatially infinite thermal system and $z,\bar{z}$ are coordinates on the vacuum system. Explicitly,
\be
\langle \mathcal{O}(x,t)...\rangle_\beta = \left(\frac{2\pi z}{\beta}\right)^h\left(\frac{2\pi \bar{z}}{\beta}\right)^{\bar{h}}\langle \mathcal{O}(z,\bar{z})...\rangle,
\ee
where $h,\bar{h}$ are the conformal weights of the $\mathcal{O}$ operator, related to the dimension and spin by $\Delta = h + \bar{h}$ and $J = h - \bar{h}$. On the left hand side, we have a thermal expectation value, at inverse temperature $\beta$, and on the right hand side we have a vacuum expectation value on the $z,\bar{z}$ space. It is common to work with units in which $\beta = 2\pi$, but we prefer to keep the $\beta$ dependence explicit.

It will be essential in this paper to study correlation functions with operators at complexified times $t_i$. In our convention, real $t$ corresponds to Minkowski time, and imaginary $t$ corresponds to Euclidean time. Notice from (\ref{z}) that $\bar{z}$ is the complex conjugate of $z$ only if the time $t$ is purely Euclidean.  In order to make contact with standard CFT formulas for the four point function, we will begin with a purely Euclidean arrangement of the operators. This means a choice of $z_1,\bar{z}_1,...,z_4,\bar{z}_4$ with $\bar{z}_i = z_i^*$. With such a configuration, the ordering of the operators is unimportant, and global conformal invariance on the $z,\bar{z}$ plane implies that the four point function can be written
\be
\langle W(z_1,\bar{z}_1)W(z_2,\bar{z}_2)V(z_3,\bar{z}_3)V(z_4,\bar{z}_4)\rangle = \frac{1}{z_{12}^{2h_w}z_{34}^{2h_v}}\frac{1}{\bar{z}_{12}^{2\bar{h}_w}\bar{z}_{34}^{2\bar{h}_v}}f(z,\bar{z}).\label{euclid}
\ee
in terms of a function $f$ of the conformally invariant cross ratios
\be
z = \frac{z_{12}z_{34}}{z_{13}z_{24}}, \hspace{20pt} \bar{z} = \frac{\bar{z}_{12}\bar{z}_{34}}{\bar{z}_{13}\bar{z}_{24}}.\label{cross}
\ee
According to the general principles of CFT, we can expand $f$ as a sum of global conformal blocks, explicitly \cite{Dolan:2000ut}
\be
f(z,\bar{z}) = \sum_{h,\bar{h}}p(h,\bar{h}) \ z^h\bar{z}^{\bar{h}} \ F(h,h,2h,z)F(\bar{h},\bar{h},2\bar{h},\bar{z})\label{blocks}
\ee
where $F$ is the Gauss hypergeometric function, the sum is over the dimensions of global $SL(2)$ primary operators, and the constants $p$ are related to OPE coefficients $p(h,\bar{h}) = \lambda_{WW\mathcal{O}_{h,\bar{h}}}\lambda_{VV\mathcal{O}_{h,\bar{h}}}$. 

\subsection{Continuation to the second sheet}\label{S3}
In order to apply the above formulas to the correlators (\ref{WVWV}) and (\ref{WWVV}), we need to understand how to obtain them as analytic continuations of the Euclidean four point function. That this is possible follows from the fact that all Wightman functions are analytic continuations of each other.\footnote{Theorem 3.6 of \cite{wightman}.} The procedure involves three steps: first, one starts with the Euclidean function, assigning small and different imaginary times $t_j = i \epsilon_j$ to each of the operators. Second, with the imaginary times held fixed, one increases the real times of the operators to the desired Lorentzian values. Finally, one smears the operators in real time and then takes the imaginary times $\{\epsilon_i\}$ to zero.\footnote{In fact, we will omit this final step in this paper. However, we will omit it consistently on both sides of the bulk and boundary calculations that we are comparing.} The result will be a Lorentzian correlator ordered such that the leftmost operator corresponds to the smallest value of $\epsilon$, the second operator corresponds to the second smallest, and so on.

This elaborate procedure is necessary because Eq.~(\ref{euclid}) is a multivalued function of the independent complex variables $\{z_i,\bar{z}_i\}$. The interesting multivaluedness (for our purposes) comes from $f(z,\bar{z})$. By crossing symmetry, this function is single valued on the Euclidean section $\bar{z} = z^*$, but it is multivalued as a function of independent $z$ and $\bar{z}$, with branch cuts extending from one to infinity. Different orderings of the $W,V$ operators correspond to different sheets of this function. To determine the correct sheet, we must assign $i\epsilon$'s as above, and follow the path of the cross ratios, watching to see if they pass around the branch loci at $z = 1$ and $\bar{z} = 1$.

To carry this out directly, we write
\begin{align}
z_1 &= e^{\frac{2\pi}{\beta}(t' + i \epsilon_1)}\hspace{20pt}\bar{z}_1 = e^{-\frac{2\pi}{\beta}(t' + i\epsilon_1)} \label{z1-config} \\
z_2 &= e^{\frac{2\pi}{\beta}(t' + i \epsilon_2)}\hspace{20pt}\bar{z}_2 = e^{-\frac{2\pi}{\beta}(t' + i\epsilon_2)}\\
z_3 &= e^{\frac{2\pi}{\beta}(x + i \epsilon_3)}\hspace{20pt}\bar{z}_3 = e^{\frac{2\pi}{\beta}(x - i\epsilon_3)}\\
z_4 &= e^{\frac{2\pi}{\beta}(x + i \epsilon_4)}\hspace{20pt}\bar{z}_4 = e^{\frac{2\pi}{\beta}(x - i\epsilon_4)} \label{z4-config}
\end{align}
as a function of the continuation parameter $t'$. When $t' = 0$, we have a purely Euclidean correlator, on the principal sheet of the function $f(z,\bar{z})$. When $t' = t>x$, we have an arrangement of operators as shown in Fig.~\ref{figtwo}.
\begin{figure}
\begin{center}
\includegraphics[scale=.5]{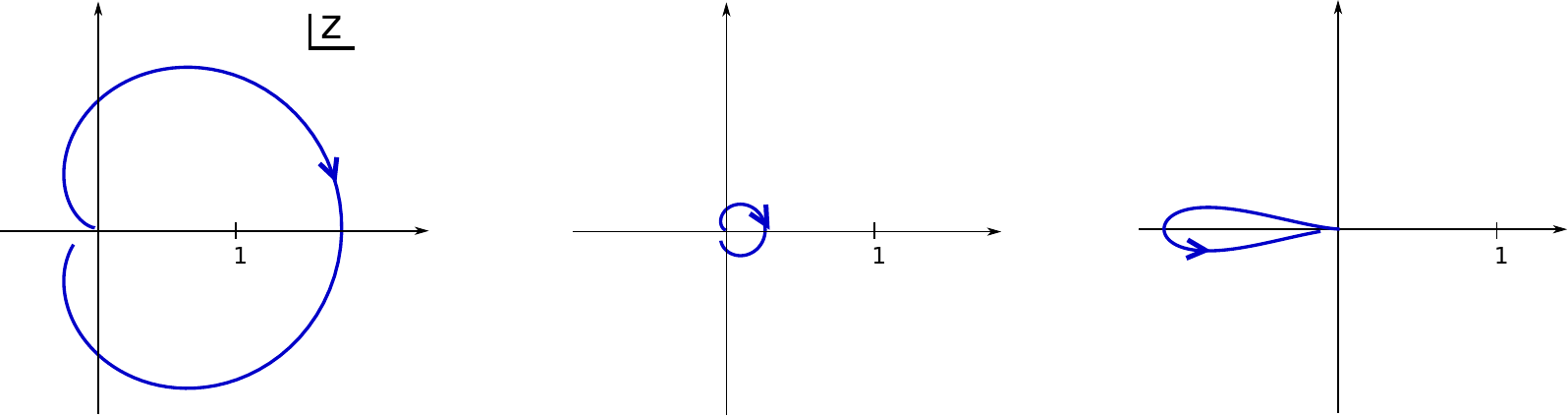}
\caption{The paths taken by the cross ratio $z$ during the continuations corresponding to (from left to right) $\langle WVWV\rangle$, $\langle WWVV\rangle$, and $\langle WVVW\rangle$. Only in the first case does the path pass around the branch point at $z = 1$. Our choice of $t >x>0$ breaks the symmetry between $z$ and $\bar{z}$, and the corresponding figures for $\bar{z}$ are very boring: the $\bar{z}$ coordinate never circles one.}\label{figone}
\end{center}
\end{figure}

The cross ratios $z,\bar{z}$ are determined by these coordinates as in Eq.~(\ref{cross}). Their paths, as a function of $t'$, depend on the ordering of operators through the associated $i\epsilon$ prescription.  Representative paths for the three cases of interest are shown in Fig.~\ref{figone}. The variable $\bar{z}$ never passes around the branch point at one, and the $z$ variable does so only in the case corresponding to $WVWV$.

In the final configuration with $t' = t$, the cross ratios are small. For $t \gg x$, we have
\be
z \approx -e^{\frac{2\pi}{\beta}(x-t)}\epsilon_{12}^*\epsilon_{34} \hspace{20pt} \bar{z} \approx -e^{-\frac{2\pi}{\beta}(x+t)}\epsilon_{12}^*\epsilon_{34}, \label{conformal-mapping}
\ee
where we introduced the abbreviation
\be
\epsilon_{ij} = i(e^{\frac{2\pi}{\beta}i\epsilon_i} - e^{\frac{2\pi}{\beta}i\epsilon_j}).
\ee
 For the orderings $WWVV$ and $WVVW$, no branch cuts are crossed, so the limit of small cross ratios can be taken on the principal sheet of Eq.~(\ref{blocks}). The contribution from the identity operator dominates, verifying our statement in the Introduction that both $\langle W(t)V VW(t)\rangle_\beta$ and $\langle W(t) W(t) V V\rangle_\beta$ approach $\langle WW\rangle\langle VV\rangle_\beta$ for large $t$.

The case corresponding to $WVWV$ is more interesting. Here, $z$ passes around the branch point at one. The hypergeometric function $F(a,b,c,z)$ has known monodromy around $z = 1$, returning to a multiple of itself, plus a multiple of the other linearly independent solution to the hypergeometric equation, $z^{1-c}F(1+a-c,1+b-c,2-c,z)$. For small $z,\bar{z}$ we then have
\be
f(z,\bar{z}) \approx \sum_{h,\bar{h}}\tilde{p}(h,\bar{h})z^{1-h}\bar{z}^{\bar{h}},\label{lorentzian}
\ee
where $\tilde{p}$ has been defined to absorb the transformation coefficient. On this sheet, as $z,\bar{z}$ become small, global primaries with large spin become important. As a function of $x,t$, individual terms in this sum grow like $e^{(h-\bar{h}-1)t}e^{-(h+\bar{h}-1)x}$. For sufficiently large $t$, this sum diverges, and it must be defind by analytic continuation. In other words, we must do the sum over $h,\bar{h}$ before we continue the cross ratios. In a CFT dual to string theory in AdS${}_3$, we expect this divergence even at a fixed order in the large $c$ expansion, because of the sum over higher spin bulk exchanges. The implications of this stringy physics for boundary CFT chaos is under investigation in \cite{stringy}.

\subsection{The Virasoro identity block}\label{S4}
We will make a few remarks about the inclusion of other operators in the Discussion. However, the primary focus of this paper is to reproduce the Einstein gravity calculation of the correlation function. In the bulk, this calculation is done by studying free propagation on a shock wave background, which implicitly sums an infinite tower of ladder exchange diagrams. In the CFT, these diagrams are related to terms involving powers and derivatives of the stress tensor in the OPE representation of the four point function. In a two dimensional CFT, all such terms can be treated simultaneously using the Virasoro conformal block of the identity operator, itself an infinite sum of $SL(2)$ conformal blocks. Including these terms (and only these!) in the OPE amounts to replacing 
\be
f(z,\bar{z})\rightarrow \mathcal{F}(z)\bar{\mathcal{F}}(\bar{z}).\label{replacement}
\ee where $\mathcal{F}$ is the Virasoro conformal block with dimension zero in the intermediate channel.

The function $\mathcal{F}$ is not known explicitly, but there are several methods for approximating it. We will use a formula from \cite{Fitzpatrick:2014vua}, which is valid at large $c$, with $h_w/c$ fixed and small, and $h_v$ fixed and large. In our notation, the formula reads
\be
\mathcal{F}(z) \approx \left(\frac{z}{1 - (1-z)^{1 - 12h_w/c}}\right)^{2h_{v}}\label{kaplan}.
\ee
This function has a branch point at $z =1$, as expected. Following the contour around $z = 1$ and taking $z$ small, we find
\be
\mathcal{F}(z)\approx \left(\frac{1}{1 - \frac{24\pi i h_w}{cz}}\right)^{2h_{v}}.\label{secondsheetF}
\ee
The trajectory of $\bar{z}$ does not circle the branch point at $\bar{z} = 1$, so for small $\bar{z}$, we simply have $\bar{\mathcal{F}}(\bar{z})\approx 1$, the contribution of the identity operator itself. Substituting (\ref{secondsheetF}) in (\ref{replacement}) and then in (\ref{euclid}), we find
\be
\frac{\langle W(t+i\epsilon_1)V(i\epsilon_3) W(t+i\epsilon_2)V(i\epsilon_4)\rangle_{\beta}}{\langle W(i\epsilon_1)W(i\epsilon_2)\rangle_\beta \ \langle V(i\epsilon_3)V(i\epsilon_4)\rangle_\beta} \approx \left(\frac{1}{1 + \frac{24\pi i\,h_w}{\epsilon_{12}^*\epsilon_{34}} e^{\frac{2\pi}{\beta}(t - t_*-x)}}\right)^{2h_{v}} \label{cft-correlator}
\ee
where we define the scrambling time $t_*$ with the convention
\be
t_* = \frac{\beta}{2\pi}\log c. \label{scrambling-time-cft}
\ee
Eq.~(\ref{cft-correlator}) is the main result of our paper. The essential point is that as $t$ increases beyond $t_* + x$, the correlation function starts to decrease, in keeping with expectations from \cite{Shenker:2013pqa,Shenker:2013yza,Roberts:2014isa}. In Appendix \ref{bulk}, we calculate the above correlation function using the gravitational shock wave techniques of the above-cited papers, finding precise agreement. We will finish this section with several comments about this formula and extensions.

\subsection{Various comments}
\subsubsection*{The $\epsilon_{12}\epsilon_{34}$ factors}
If we take the $\epsilon_j$ parameters to zero, the denominator in (\ref{scrambling-time-cft}) diverges. This because, in this limit, we have two pairs of coincident operators on the LHS, and the expectation value is dominated by the high energy components of the operators. The function tends to zero because the high energy components disturb the state violently and decorrelate faster. We can avoid this divergence by smearing $W$ and $V$ over an interval of Lorentzian time $L$ before taking the $\epsilon$ parameters to zero. This procedure will replace the $\epsilon_{12}\epsilon_{34}$ factors with $\sim L^2$. Instead of doing this explicitly, we will leave the $\{\epsilon_j\}$ finite. This is another way of cutting off the high energy components, essentially inserting $e^{-\epsilon H}$ between each operator. Either way, $\epsilon_{12}\epsilon_{34}h_wh_v$ should be understood as proportional to $\beta^2 E_w E_v$, the product of the energy scales of the operators.

\subsubsection*{Validity of the block formula}
The derivation of (\ref{kaplan}) in \cite{Fitzpatrick:2014vua} is valid for large but fixed $h_v$, in the limit of large $c$ with $h_w/c$ fixed and small. It is accurate uniformly in $z$, including on the second sheet. With this scaling, $W$ is a heavy operator. In the context of quantum chaos, we are interested in a somewhat different limit, with light operators and long times: $h_w,h_v$, and $cz$ fixed (on the second sheet) as $c\rightarrow \infty$. The distinction is subtle but important. With $h_w/c$ fixed, the correlation function (\ref{cft-correlator}) becomes affected at a time $t \sim \frac{\beta}{2\pi}\log c/h_w$, which is order one as a function of $c$. On the other hand, with $h_w$ fixed, this timescale grows with $c$. 

Based on agreement with bulk calculations, it seems that formula (\ref{cft-correlator}) is actually still valid in this scaling, provided that $h_w\gg h_v\gg 1$. It would be good to derive this fact directly. In the meantime, we note the $SL(2)$ blocks are enough to show that (for fixed $h_w,h_v$, large $c$) the time until the correlator is affected is of order $t_*= \frac{\beta}{2\pi}\log c$. This criterion comes from the contribution of the stress tensor to (\ref{lorentzian}), which is proportional to $h_vh_wz/c \sim h_vh_we^{\frac{2\pi}{\beta}(t-t_*-x)}$.

\subsubsection*{Convergence in $1/c$}
Notice that the power series in $1/c$ diverges when the second term in the denominator becomes equal to one. This is for $t\approx x + t_*$. However, the correlation function is perfectly well behaved at that point. This illustrates the importance of doing the sum over $h,\bar{h}$ before continuing onto the second sheet.

\subsubsection*{Asymmetry in $h_w$ and $h_v$}
The expression on the LHS is basically symmetric under interchanging $W$ and $V$, but our answer on the RHS is definitely not. This is because we have assumed that $h_w \gg h_v$. We assumed this for purely technical reasons---the Virasoro identity block is not known in the limit of interest if $h_w$ and $h_v$ are comparable. Using holography, we can write the answer for general $h_w,h_v$ as an integral over momenta and transverse coordinates. We give this expression in (\ref{integral}) of appendix (\ref{bulk}). This constitutes a prediction for the behavior of the Virasoro identity block at small values of $z$ on the second sheet. It would be interesting to check this directly in conformal field theory.

\subsubsection*{A two-sided configuration}
Although the correlation function (\ref{cft-correlator}) is a purely `one-sided' quantity, defined with respect to the thermal expectation value, its behavior is closely related to the disruption of `two-sided' entanglement of the thermofield double state, as studied in \cite{Shenker:2013pqa,Shenker:2013yza,Leichenauer:2014nxa}.

The thermofield double state is a purification of the thermal ensemble we have been considering, given by
\be
|TFD\rangle = \frac{1}{\sqrt{Z(\beta)}}\sum_i e^{-\beta E_i/2} |i\rangle_L |i\rangle_R,
\ee
where two noninteracting copies of the system are denoted $L$ and $R$, and $Z(\beta)$ is the thermal partition function of an individual system. Given an operator $O$, we define $O_R = 1\otimes O$ acting on the $R$ system and $O_L = O^T\otimes 1$ acting on the $L$ system. The thermofield double state has a high degree of local correlation between these pairs. As shown in \cite{Shenker:2013pqa}, a small perturbation to one of the systems at an early time will disrupt the local entanglement. We can diagnose this by studying correlation functions in the perturbed state $|\psi \rangle = W(t)_R |TFD\rangle$
\be
\langle \psi | V_L V_R|\psi \rangle = \langle TFD | W(t)_R V_L V_R W(t)_R | TFD \rangle.
\ee 
It is easy to check that this correlation function is related to a one-sided correlation function of the type (\ref{WVWV}) by analytic continuation,
\be
\langle \psi | V_L V_R  |\psi \rangle = \langle W(t) V W(t) V(i \beta/2 ) \rangle_\beta.
\ee
We can make this continuation directly in (\ref{cft-correlator}) by setting $\epsilon_4 = \beta/2$ and $\epsilon_3=0$. Then $\epsilon_{34} \rightarrow 2i$. Again, as the time $t$ increases past $t_* + x$, the two-sided correlation begins to decrease exponentially.

\subsubsection*{Commutators and growth of operators}
The correlation function (\ref{cft-correlator}) is closely related to the square of the commutator
\be
\langle [W(t),V]^2\rangle_\beta.
\ee
As the correlation function (\ref{cft-correlator}) becomes small, the commutator becomes large, of order $\langle WW\rangle\langle VV\rangle$. In \cite{Roberts:2014isa}, the behavior of this commutator was used to determine the growth of a precursor operator such as $W(t)$. Here, we think about $V$ as a probe to determine the size of $W(t)$. Translated in terms of the correlator (\ref{cft-correlator}), the definition of \cite{Roberts:2014isa} is that the size of $W(t)$ is the volume of the region (as a function of the location of $V$) in which the correlator is less than one half. This volume is a ball, and radius of the operator $r[W(t)]$ is defined as the radius of that ball. For $d$ dimensional CFTs dual to Einstein gravity, \cite{Shenker:2013pqa,Roberts:2014isa} found that the ball has radius
\be
r[W(t)] \approx v_B(t-t_*), \hspace{20pt} v_B = \sqrt{\frac{d}{2(d-1)}}. \label{cft-precursor-growth}
\ee
Specializing to $d = 2$, we find $v_B = 1$, in agreement with (\ref{cft-correlator}). 

\subsubsection*{Two-sided entropy}
Another two-sided quantity of interest is the entropy $S_{AB}$ between two matching semi-infinite intervals, one on each boundary. Apart from the usual UV divergence near the boundary of the intervals, this entropy is finite. In other words, there is no IR divergence despite the infinite length of the intervals. As pointed out in \cite{Hartman:2013qma}, this is related to the local entanglement of the two sides of the thermofield state.

As discussed above, applying a small local perturbation at an early time will disrupt this entanglement and cause the entropy $S_{AB}$ to grow. Using the gravity techniques of \cite{Shenker:2013pqa,Roberts:2014isa}, one can check that the mutual information decreases linearly with the time of this perturbation, $t$, as (for large $t$)
\be
\Delta S_{AB} = \frac{\pi c}{3\beta} (|t| - t_* -  |x| ) + (const),\label{mutual-information}
\ee
where the constant depends on the smearing and dimension of the perturbing operator, and where $x$ is the distance outside the interval that the perturbation was applied. In this paper, we chose to focus on four point functions instead of entropy, but the Virasoro identity block can also be used to calculate the above, following very closely the techniques in \cite{Asplund:2014coa} and the conformal mappings used in this section. We have verified that the answer matches.\footnote{A similar problem was recently studied by \cite{Caputa:2014eta}, and a linear increase in the entropy was found, but the formula differs significantly from (\ref{mutual-information}).}

\subsubsection*{Scrambling in the vacuum}
In this paper, we have focused on thermal expectation values. It is interesting to compare to similar correlation functions, evaluated in the vacuum. Taking limit of \eqref{z1-config}--\eqref{z4-config} for infinite $\beta$, the conformal cross ratio for $t \gg x$ becomes $z \approx  -(\epsilon_1 - \epsilon_2)(\epsilon_3 - \epsilon_4)/(t-x)^2$. In this configuration, the contour still crosses the branch cut, and we find
\be
\langle WVWV \rangle \propto  \left(\frac{1}{1 + \frac{24\pi i\,h_w (t-x)^2}{c (\epsilon_1 - \epsilon_2)(\epsilon_3 - \epsilon_4) } }\right)^{2h_{v}},
\ee
where in this case the expectation is taken in the vacuum state.

We can understand this result as a type of ``slow scrambling'' that happens in an extremely low temperature state on an infinite line, with $\beta \gg \sqrt{c} L$, where $L$ is the smearing scale of the operators. In that case, for $L\sqrt{c}<t<\beta$, the correlation will decrease with power law dependence as $\sim t^{-4 h_v}$. After $t$ becomes comparable to $\beta$, the dependence will be given by \eqref{cft-correlator} and will decrease exponentially.

\section{Discussion}

Several manifestations of chaos in quantum field theory have recently been discussed: disruption of atypical correlations by small perturbations, \cite{Shenker:2013pqa, Leichenauer:2014nxa}, large commutators \cite{Shenker:2013yza,kitaev}, linear growth of precursor operators \cite{Roberts:2014isa}, ---all of which are ultimately connected to an out-of-time-order correlation function of the form $\langle W(t)VW(t)V \rangle_\beta$. Previous studies have relied on holography. In this paper, we showed how to obtain the relevant correlation function starting from the Euclidean correlator, following \cite{Cornalba:2006xk} and \cite{Hartman:2013mia}. We then used an approximation to the Euclidean correlation function, the Virasoro identity block \cite{Hartman:2013mia,Fitzpatrick:2014vua}, to compute the out-of-time order correlation function. We found agreement with a generalization of the original holographic calculation of \cite{Shenker:2013pqa}.

However, it is clear both from the perspective of holography and conformal field theory that there is more to the correlation function than our main result, \eqref{cft-correlator}. In the bulk, we see this in the form of stringy corrections to the correlator (which are important at large $t$) \cite{stringy}. In conformal field theory, we have to consider the effect of operators that are not Virasoro descendants of the identity. We can make some progress from the CFT side by considering additional operators in the channel of the four-point function. 

To explore these contributions, let us compare the second-sheet contribution of the stress tensor, $\propto e^{t-x}$, to the contribution of a general $SL(2)$ primary, $e^{(h-\bar{h}-1)t - (h+\bar{h}-1)x}$. At leading order in $1/c$, the most important contributions should be from single-trace operators. In a theory with a weakly curved holographic dual, the lowest-dimension single trace operators with spin greater than two should have parametrically large dimension. A useful example to keep in mind is a low $n$ string state on the leading Regge trajectory, with $h - \bar{h} = n$ and $h+\bar{h}\sim \sqrt{n}\, m_s\ell_{AdS}$. The contribution of this type of operator is suppressed as a function of $x$, but at sufficiently large $t$, such that
\be
 (h - \bar{h} - 2)t >(h + \bar{h} -2)x,\label{criterion}
\ee
it will dominate over the stress tensor. If we fix $x$, then at $t_*$, the above criterion will be satisfied. This is in keeping with the bulk analysis of \cite{stringy}, which shows that a Regge-type resummation of an infinite number of stringy operators is necessary.

At sufficiently large $x$, this criterion will not be satisfied (here, we are assuming large $m_s\ell_{AdS}$), even at the relevant scrambling time $t = t_*+x$. This indicates a region in which identity block will dominate. We would like to suggest that there is another: very large $t \gg t_*+x$. Our reason for this suspicion is the bulk geodesic `bending' discussed in appendix \ref{bulk}. For simplicity, let us set $x \approx 0$. Then for $t>t_*$, the important correlation comes from bulk geodesics that pass through the horizon at transverse coordinate $|x'| \sim (t-t_*)/2$. If this $x'$ is large enough to violate (\ref{criterion}), we expect the extra operator to be subleading compared to the identity. 

To put a little flesh on this idea, let us consider the Virasoro block of the $h,\bar{h}$ operator, rather than the $SL(2)$ block. Using, again, the results from \cite{Fitzpatrick:2014vua} and continuting to the second sheet, we find
\be
\frac{\mathcal{F}_h(z)}{\mathcal{F}(z)} \approx \left( \frac{16}{z- \frac{24\pi i h_w}{c} }  \right)^{h}
\ee
where $\mathcal{F}(z)$ is the identity block \eqref{cft-correlator}. For $x = 0$ and $t > t_*$, this ratio is proportional to $e^{\frac{2\pi}{\beta}h t_*}$. Together with the contribution from the antiholomorphic block, which is $e^{-\frac{2\pi}{\beta}\bar{h}t}$, we find that the $h,\bar{h}$ operator will dominate only if
\be
\bar{h} t < h t_*.
\ee
For large $E=h+\bar{h}$ and to leading order in $J/E = (h - \bar{h})/(h+\bar{h})$, this is consistent with the criterion (\ref{criterion}) for the $SL(2)$ blocks, but evaluated at the `bent' location $x' = (t - t_*)/2$.

Finally, we will comment on two possible extensions that were not indicated above. First, using methods outlined in \cite{Hartman:2013mia}, it seems possible to extend our analysis to multi-point functions dual to the multiple shock wave backgrounds studied in \cite{Shenker:2013yza,Roberts:2014isa}. Second, we note that shock wave geometries and de-correlation have been studied in relation to tensor networks \cite{Stanford:2014jda} and computational complexity \cite{Susskind:2014rva,Susskind:2014ira,Susskind:2014jwa,Susskind:2014moa}. It would be interesting to translate the Virasoro identity block approximation (and corrections) into the language of those papers.

\section*{Acknowledgments}
We are grateful to Ethan Dyer, Tom Hartman, Alexei Kitaev, and Steve Shenker for helpful discussions. D.R. is supported by the Fannie and John Hertz Foundation and is very thankful for the hospitality of the Stanford Institute for Theoretical Physics during the completion of this work. D.R. also acknowledges the U.S. Department of Energy under cooperative research agreement Contract Number  DE-SC00012567. D.S. is supported by NSF grant PHY\_1314311/Dirac. This paper has been brought to you by the letters $W$, $\bar{z}$, and $c$, and the number $i$.

\appendix

\section{Bulk calculations}\label{bulk}

In this section we will compute the correlation function (\ref{cft-correlator}) using the gravitational shock wave methods of \cite{Shenker:2013pqa,Roberts:2014isa}. The idea is as follows. If $h_w \gg h_v\gg 1$, we can calculate the correlation function by treating the $W$ operator as creating a shock wave, and calculating the two point function of the $V$ operator on that background. Although the calculation can be done for general $\epsilon_j$, we will focus on the case where $\epsilon_1 = -\tau$, $\epsilon_2 = \tau$, $\epsilon_3 = 0$ and $\epsilon_4 = \beta/2$, because this corresponds to a two-sided expectation value
\be
\frac{\langle \psi|V_L V_R|\psi\rangle}{\langle \psi|\psi\rangle\langle V_L V_R\rangle}\label{yoyo}
\ee
in the state
\be
|\psi\rangle = W_R(t + i\tau)|TFD\rangle.
\ee
Here, the normalization factor $\langle V_L V_R\rangle$ is evaluated in the unperturbed thermofield double state. Eq.~(\ref{yoyo}) is the correlation function studied in \cite{Shenker:2013pqa}. The analysis breaks into two parts: {\it (i)} finding the geometry of the shock sourced by $W$ and {\it (ii)} computing the correlation function of the $V$ operators in that background.

\subsection{The shock geometry}
The metric of a localized shock wave\footnote{Although we refer to these geometries as shock waves, the terminology is somewhat misleading in 2+1 dimensions, since the geometry is locally pure AdS${}_3$ away from the source.} in 2+1 dimensional AdS Rindler space, is \cite{Dray:1984ha,Horowitz:1999gf,Cornalba:2006xk,Shenker:2013pqa,Roberts:2014isa}
\begin{align}
ds^2 &= -\frac{4}{(1+uv)^2} dudv + \frac{(1-uv)^2}{(1+uv)^2} dx^2 + 4 \delta(u) h(x) du^2, \label{metric}
\end{align}
where $h$ will be defined below. Schwarzschild $(r,t)$ are related to the Kruskal $(u,v)$ by 
\be
r= \frac{1-uv}{1+uv}, \qquad t = \frac{1}{2} \log \Big( -\frac{v}{u} \Big).
\ee
The inverse temperature $\beta$ is $2\pi$. We will consider a shock sourced by a stress tensor
\be
T_{uu}(u',v',x') = P \, \delta(u') \delta(x'-x), \label{stress-tensor}
\ee
appropriate for a particle sourced by the $W$ operator, traveling along the $u = 0$ horizon at transverse position $x$. The constant $P$ is related to the momentum $p^v$ of the particle. The metric (\ref{metric}) can be understood as two halves of AdS-Rindler, glued together at $u = 0$ with a shift 
\be
\delta v(x) = h(x)
\ee
in the $v$ direction. Plugging in to Einstein's equations, we determine $h$ as
\be
h(x') = 2\pi G_N P e^{-|x'-x|} \label{shift}.
\ee

To relate this geometry to the state $|\psi\rangle$, we need to fix $P$. In other words, we need to evaluate
\be
\frac{\langle\psi| \int dx' du' ~ T_{uu} | \psi \rangle}{\langle\psi| \psi \rangle} , \label{match}
\ee
where we take the integral to run over the slice $v =0$. We will assume that $W$ is dual to a single-particle operator in the bulk, so that the state $|\psi\rangle$ can be described by a Klein-Gordon wave function $K$. This wave function is a bulk-to boundary propagator from the location $(x,t)$ of the $W$ operator. It is given in terms of the regularized geodesic distance $d$ from the boundary point, as $K \propto (\cosh d)^{-{2h_w}} $. At $v = 0$, we find
\be
K(t,x; u', x') = \frac{\mathcal{N}}{[e^t u' +\cosh(x-x') ]^{2h_w}}. \label{bulk-to-boudary}
\ee
$\mathcal{N}$ is a normalization that will ultimately cancel, so we will set it to one.

We will evaluate the denominator of \eqref{match} first. The norm $\langle\psi| \psi \rangle$ is a Klein-Gordon inner product
\be
\langle\psi| \psi \rangle = 2i \int dx' du' ~K(t + i \tau, x; u', x')^* \partial_{u'} K(t + i \tau, x; u', x').
\ee
The $u'$ integral can be done using contour integration, and the $x'$ integral can be done in terms of $\Gamma$ functions:
\be
\langle\psi| \psi \rangle = \frac{4 \pi^{3/2} }{(2 \sin \tau)^{{4h_w}}} \frac{\Gamma[{4h_w}]}{\Gamma[{2h_w}] \Gamma[{2h_w}+\frac{1}{2}]}.
\ee

For the numerator, the stress tensor for the Klein-Gordon field is given by the expression $T_{uu}= \partial_u \varphi \partial_u \varphi$. Contracting bulk operators with boundary operators using $K$, we have 
\be
\langle\psi| \int dx' du' ~ T_{uu}| \psi \rangle= 2\int dx' du' ~\partial_{u'}  K(t_w + i \tau, x; u', x')^* \partial_{u'} K(t_w + i \tau, x; u', x'),
\ee
where the factor of $2$ comes from the two different ways of doing the contractions. The integrals can be done the same way as before:
\be
\langle\psi| \int dx' du' ~ T_{uu}| \psi \rangle= \frac{8 \pi^{3/2}  e^{t_w}}{(2\sin \tau)^{{4h_w}+1}} \frac{\Gamma[4{h_w}]\Gamma[{2h_w}+\frac{1}{2}]}{\Gamma[{2h_w}]^3}.
\ee
Taking the ratio at large ${h_w}$, we find
\be
P =  \frac{{2h_w} \,e^t}{\sin \tau}. \label{energy-of-shock}
\ee

\subsection{The two point correlator}

The second step, following \cite{Shenker:2013pqa}, is to compute the two-sided correlation function of the $V$ operators in this shock background. We will do this using the geodesic approximation 
\be
\langle\psi| V_L V_R |\psi\rangle \propto e^{-m d}, \label{geodesic-approximation}
\ee
where $d$ is the regularized geodesic distance, and the mass $m$ is approximately $2h_v$, the conformal weight of $V$.

The relevant geodesic passes from $t=0$ and $x = 0$ on the left boundary to $t = 0$ and $x = 0$ on the right. As in \cite{Shenker:2013pqa}, we first compute the length of a geodesic from the left boundary to an intermediate point on the $u=0$ surface $(v',x')$, and then we add it to the length from that point to the right boundary. These distances can be worked out from the embedding coordinates:
\be
d = 2 \log 2 r_\infty + \log \Big[ \cosh x' - v' \Big] + \log \Big[ \cosh x' + v' + h(x') \Big],
\ee
where $r_\infty$ is the cutoff at the boundary of AdS, and $h(x)$ is given by \eqref{shift}. The true geodesic distance is given by extremizing this sum over the intermediate point. This gives
\be
d = 2 \log 2 r_\infty + \log \Big[ 1 + h(0) \Big]. \label{distance}
\ee
Note the difference from the behavior $2\log(1 + h/2)$ in the spherically symmetric setting of \cite{Shenker:2013pqa}. At large $h(0)$, the distance is half of what one might naively expect based on the spherical problem. The reason is that, for $t > t_* +x$, the geodesic bends away from the source of the shock, passing through at transverse position
\be
x'\approx \frac{t_*+x-t}{2}<0,
\ee
here $x>0$ is the transverse position of the source of the shock. 

After subtracting the divergent distance in the unperturbed thermofield double state, $d_{TFD} = 2\log 2 r_{\infty}$, we plug the distance into (\ref{geodesic-approximation}), finding
\be
\frac{\langle \psi|V_L V_R|\psi\rangle}{\langle \psi|\psi\rangle\langle V_L V_R\rangle} = \Bigg( \frac{1}{1+\frac{4\pi G_N h_w}{\sin \tau}e^{t-x}}\Bigg)^{2h_v}.
\ee
This agrees with (\ref{cft-correlator}) after {\it (i)} using $G_N = 3/2c$ to express the gravitational constant in terms of the central charge, {\it(ii)} plugging in $\epsilon_1 = -\tau$, $\epsilon_2 = \tau$, $\epsilon_3 = 0$, $\epsilon_4 = \beta/2$ and {\it (iii)} using $\beta = 2\pi$.

\subsection{Generalization}
The case with general $\epsilon_j$ and general $h_w,h_v$ is analyzed in \cite{stringy}. The calculation is similar to the above, but includes an integral over the momentum of the shock, weighted by wave functions associated to the $W$ operator. We will not give the details of this calculation here, but we will note that it can be used to give a prediction for the behavior of the Virasoro conformal block $\mathcal{F}$
\be
\mathcal{F}(z) = c_0^2\int_0^\infty dp\,dq\int_{-\infty}^\infty dx\, dy \ p^{4h_w-1}q^{4h_v-1}e^{-p\cosh x}e^{-q\cosh y}e^{3\pi i pq \,e^{x+y}/cz}.\label{integral}
\ee
where the normalization constant ensures that $\mathcal{F}(z) = 1$ when $c = \infty$:
\be
c_0 = \frac{2^{1-2h_w-2h_v}}{\Gamma[2h_w]\Gamma[2h_v]}.
\ee
We remind the reader that this expression is on the second sheet, where $z$ has passed around the branch point at $z = 1$. For $h_w\gg h_v\gg 1$, one can check that this reduces to (\ref{cft-correlator}). However, the formula should be accurate for all $h_w,h_v$ in the large $c$ limit with $cz$ fixed. It would be interesting to check this directly in CFT.

\section{Ising model} \label{ising}
It is interesting to contrast the behavior of out-of-time-order correlators in a chaotic theory to those in an integrable theory. In a chaotic theory, we expect all thermal correlators of the form $\langle W(t) V W(t) V \rangle_\beta$ should become small for sufficiently large $t$, regardless of which operators $V$ and $W$ are correlated. This will not be the case in an integrable theory. 

As an example, we consider the two-dimensional Ising CFT. This theory has $c=1/2$ and three Virasoro primary operators: $I$, $\sigma$, and $\epsilon$, corresponding to the identity, `spin,' and `energy' operators. The different combinations of four-point correlators of these primaries are well known \cite{Belavin:1984vu,Mattis:1986mj,Ginsparg:1988ui}. We will present these by giving the function $f(z,\bar{z})$ in Eq.(\ref{euclid}) for the three cases: 
\begin{align}
f_{\sigma\sigma}(z,\bar{z}) &= \frac{1}{2}\bigg| \frac{1}{1-z}\bigg|^{1/4}  \Big( \big|1+\sqrt{1-z}\big| + \big|1-\sqrt{1-z}\big| \Big), \\
f_{\sigma\epsilon}(z,\bar{z}) &= \bigg| \frac{2-z}{2\sqrt{1-z}} \bigg|^2, \\
f_{\epsilon\epsilon}(z,\bar{z}) &= \bigg| \frac{1-z+z^2}{1-z} \bigg|^2,
\end{align}
where the subscripts in $f_{WV}$ denote $W$ and $V$, and the operators are ordered $WVWV$ with the configuration specified by \eqref{z1-config}--\eqref{z4-config}.\footnote{{\bf[v3]:} We thank Raghu Mahajan for pointing out an error in the prefactor of $f_{\sigma\sigma}$. The overall conclusion is unchanged.} $z$ and $\bar{z}$ still given by \eqref{conformal-mapping}. Since we treat $z,\bar{z}$ as independent, the notation $|h(z)|$ should be interpreted as $\sqrt{h(z)}\sqrt{h(\bar{z})}$. It's easy to see that the $\langle \sigma \sigma \sigma \sigma \rangle$ and $\langle \sigma \epsilon \sigma \epsilon \rangle$ correlators have branch points at $z=1$, while $\langle \epsilon \epsilon \epsilon \epsilon \rangle$ does not. Following the contour across the branch cut for the two correlators that do have a second sheet and taking $z$ small, we find 
\be
\frac{\langle \sigma \sigma \sigma\sigma \rangle_\beta}{\langle \sigma \sigma \rangle_\beta^2} = 0, \qquad \frac{\langle \sigma \epsilon \sigma \epsilon \rangle_\beta}{\langle \sigma \sigma \rangle_\beta \langle \epsilon \epsilon \rangle_\beta} = -1, \qquad \frac{\langle \epsilon \epsilon \epsilon\epsilon \rangle_\beta}{ \langle \epsilon \epsilon \rangle_\beta^2} = 1.
\ee
Only $\langle \sigma \sigma \sigma \sigma \rangle$ vanishes at large $t$.

\mciteSetMidEndSepPunct{}{\ifmciteBstWouldAddEndPunct.\else\fi}{\relax}
\bibliographystyle{utphys}
\bibliography{cft}{}

\ifx\mcitethebibliography\mciteundefinedmacro
\PackageError{utphys.bst}{mciteplus.sty has not been loaded}
{This bibstyle requires the use of the mciteplus package.}\fi
\providecommand{\href}[2]{#2}\begingroup\raggedright\begin{mcitethebibliography}{10}

\bibitem{Shenker:2013pqa}
S.~H. Shenker and D.~Stanford, ``{Black holes and the butterfly effect},''
\href{http://arxiv.org/abs/1306.0622}{{\ttfamily arXiv:1306.0622 [hep-th]}}.

\bibitem{Shenker:2013yza}
S.~H. Shenker and D.~Stanford, ``{Multiple Shocks},''
\href{http://arxiv.org/abs/1312.3296}{{\ttfamily arXiv:1312.3296 [hep-th]}}.

\bibitem{Leichenauer:2014nxa}
S.~Leichenauer, ``{Disrupting Entanglement of Black Holes},''
  \href{http://dx.doi.org/10.1103/PhysRevD.90.046009}{{\em Phys.Rev.}
  {\bfseries D90} (2014) 046009},
\href{http://arxiv.org/abs/1405.7365}{{\ttfamily arXiv:1405.7365 [hep-th]}}.

\bibitem{Roberts:2014isa}
D.~A. Roberts, D.~Stanford, and L.~Susskind, ``{Localized shocks},''
\href{http://arxiv.org/abs/1409.8180}{{\ttfamily arXiv:1409.8180 [hep-th]}}.

\bibitem{kitaev}
A.~Kitaev, ``Hidden correlations in the hawking radiation and thermal noise.''.
  Talk given at the Fundamental Physics Prize Symposium, Nov. 10, 2014.

\bibitem{Hartman:2013mia}
T.~Hartman, ``{Entanglement Entropy at Large Central Charge},''
\href{http://arxiv.org/abs/1303.6955}{{\ttfamily arXiv:1303.6955 [hep-th]}}.

\bibitem{Fitzpatrick:2014vua}
A.~L. Fitzpatrick, J.~Kaplan, and M.~T. Walters, ``{Universality of
  Long-Distance AdS Physics from the CFT Bootstrap},''
  \href{http://dx.doi.org/10.1007/JHEP08(2014)145}{{\em JHEP} {\bfseries 1408}
  (2014) 145},
\href{http://arxiv.org/abs/1403.6829}{{\ttfamily arXiv:1403.6829 [hep-th]}}.

\bibitem{Asplund:2014coa}
C.~T. Asplund, A.~Bernamonti, F.~Galli, and T.~Hartman, ``{Holographic
  Entanglement Entropy from 2d CFT: Heavy States and Local Quenches},''
\href{http://arxiv.org/abs/1410.1392}{{\ttfamily arXiv:1410.1392 [hep-th]}}.

\bibitem{larkin}
A.~Larkin and Y.~Ovchinnikov, ``{Quasiclassical method in the theory of
  superconductivity},'' {\em JETP} {\bfseries 28,6} (1969) 1200--1205.

\bibitem{Cornalba:2006xk}
L.~Cornalba, M.~S. Costa, J.~Penedones, and R.~Schiappa, ``{Eikonal
  Approximation in AdS/CFT: From Shock Waves to Four-Point Functions},''
  \href{http://dx.doi.org/10.1088/1126-6708/2007/08/019}{{\em JHEP} {\bfseries
  0708} (2007) 019},
\href{http://arxiv.org/abs/hep-th/0611122}{{\ttfamily arXiv:hep-th/0611122
  [hep-th]}}.

\bibitem{Cornalba:2006xm}
L.~Cornalba, M.~S. Costa, J.~Penedones, and R.~Schiappa, ``{Eikonal
  Approximation in AdS/CFT: Conformal Partial Waves and Finite N Four-Point
  Functions},'' \href{http://dx.doi.org/10.1016/j.nuclphysb.2007.01.007}{{\em
  Nucl.Phys.} {\bfseries B767} (2007) 327--351},
\href{http://arxiv.org/abs/hep-th/0611123}{{\ttfamily arXiv:hep-th/0611123
  [hep-th]}}.

\bibitem{Cornalba:2007zb}
L.~Cornalba, M.~S. Costa, and J.~Penedones, ``{Eikonal approximation in
  AdS/CFT: Resumming the gravitational loop expansion},''
  \href{http://dx.doi.org/10.1088/1126-6708/2007/09/037}{{\em JHEP} {\bfseries
  0709} (2007) 037},
\href{http://arxiv.org/abs/0707.0120}{{\ttfamily arXiv:0707.0120 [hep-th]}}.

\bibitem{Hayden:2007cs}
P.~Hayden and J.~Preskill, ``{Black holes as mirrors: Quantum information in
  random subsystems},''
  \href{http://dx.doi.org/10.1088/1126-6708/2007/09/120}{{\em JHEP} {\bfseries
  0709} (2007) 120},
\href{http://arxiv.org/abs/0708.4025}{{\ttfamily arXiv:0708.4025 [hep-th]}}.

\bibitem{Sekino:2008he}
Y.~Sekino and L.~Susskind, ``{Fast Scramblers},''
  \href{http://dx.doi.org/10.1088/1126-6708/2008/10/065}{{\em JHEP} {\bfseries
  0810} (2008) 065},
\href{http://arxiv.org/abs/0808.2096}{{\ttfamily arXiv:0808.2096 [hep-th]}}.

\bibitem{Lashkari:2011yi}
N.~Lashkari, D.~Stanford, M.~Hastings, T.~Osborne, and P.~Hayden, ``{Towards
  the Fast Scrambling Conjecture},''
  \href{http://dx.doi.org/10.1007/JHEP04(2013)022}{{\em JHEP} {\bfseries 1304}
  (2013) 022},
\href{http://arxiv.org/abs/1111.6580}{{\ttfamily arXiv:1111.6580 [hep-th]}}.

\bibitem{Brown:2012gy}
W.~Brown and O.~Fawzi, ``{Scrambling speed of random quantum circuits},''
\href{http://arxiv.org/abs/1210.6644}{{\ttfamily arXiv:1210.6644 [quant-ph]}}.

\bibitem{stringy}
S.~H. Shenker and D.~Stanford, ``{Stringy effects in scrambling},''
\href{http://arxiv.org/abs/1412.6087}{{\ttfamily arXiv:1412.6087 [hep-th]}}.

\bibitem{Jackson:2014nla}
S.~Jackson, L.~McGough, and H.~Verlinde, ``{Conformal Bootstrap, Universality
  and Gravitational Scattering},''
\href{http://arxiv.org/abs/1412.5205}{{\ttfamily arXiv:1412.5205 [hep-th]}}.

\bibitem{Dolan:2000ut}
F.~Dolan and H.~Osborn, ``{Conformal four point functions and the operator
  product expansion},''
  \href{http://dx.doi.org/10.1016/S0550-3213(01)00013-X}{{\em Nucl.Phys.}
  {\bfseries B599} (2001) 459--496},
\href{http://arxiv.org/abs/hep-th/0011040}{{\ttfamily arXiv:hep-th/0011040
  [hep-th]}}.

\bibitem{wightman}
R.~F. Streater and A.~S. Wightman, {\em PCT, spin and statistics, and all
  that}.
\newblock W.A. Benjamin, Inc., 1964.

\bibitem{Hartman:2013qma}
T.~Hartman and J.~Maldacena, ``{Time Evolution of Entanglement Entropy from
  Black Hole Interiors},''
  \href{http://dx.doi.org/10.1007/JHEP05(2013)014}{{\em JHEP} {\bfseries 1305}
  (2013) 014},
\href{http://arxiv.org/abs/1303.1080}{{\ttfamily arXiv:1303.1080 [hep-th]}}.

\bibitem{Caputa:2014eta}
P.~Caputa, J.~Simon, A.~Stikonas, and T.~Takayanagi, ``{Quantum Entanglement of
  Localized Excited States at Finite Temperature},''
\href{http://arxiv.org/abs/1410.2287}{{\ttfamily arXiv:1410.2287 [hep-th]}}.

\bibitem{Stanford:2014jda}
D.~Stanford and L.~Susskind, ``{Complexity and Shock Wave Geometries},''
  \href{http://dx.doi.org/10.1103/PhysRevD.90.126007}{{\em Phys.Rev.}
  {\bfseries D90} (2014) 126007},
\href{http://arxiv.org/abs/1406.2678}{{\ttfamily arXiv:1406.2678 [hep-th]}}.

\bibitem{Susskind:2014rva}
L.~Susskind, ``{Computational Complexity and Black Hole Horizons},''
\href{http://arxiv.org/abs/1402.5674}{{\ttfamily arXiv:1402.5674 [hep-th]}}.

\bibitem{Susskind:2014ira}
L.~Susskind, ``{Addendum to Computational Complexity and Black Hole
  Horizons},''
\href{http://arxiv.org/abs/1403.5695}{{\ttfamily arXiv:1403.5695 [hep-th]}}.

\bibitem{Susskind:2014jwa}
L.~Susskind and Y.~Zhao, ``{Switchbacks and the Bridge to Nowhere},''
\href{http://arxiv.org/abs/1408.2823}{{\ttfamily arXiv:1408.2823 [hep-th]}}.

\bibitem{Susskind:2014moa}
L.~Susskind, ``{Entanglement is not Enough},''
\href{http://arxiv.org/abs/1411.0690}{{\ttfamily arXiv:1411.0690 [hep-th]}}.

\bibitem{Dray:1984ha}
T.~Dray and G.~'t~Hooft, ``{The Gravitational Shock Wave of a Massless
  Particle},''
\href{http://dx.doi.org/10.1016/0550-3213(85)90525-5}{{\em Nucl.Phys.}
  {\bfseries B253} (1985) 173--188}.

\bibitem{Horowitz:1999gf}
G.~T. Horowitz and N.~Itzhaki, ``{Black holes, shock waves, and causality in
  the AdS / CFT correspondence},''
  \href{http://dx.doi.org/10.1088/1126-6708/1999/02/010}{{\em JHEP} {\bfseries
  9902} (1999) 010},
\href{http://arxiv.org/abs/hep-th/9901012}{{\ttfamily arXiv:hep-th/9901012
  [hep-th]}}.

\bibitem{Belavin:1984vu}
A.~Belavin, A.~M. Polyakov, and A.~Zamolodchikov, ``{Infinite Conformal
  Symmetry in Two-Dimensional Quantum Field Theory},''
\href{http://dx.doi.org/10.1016/0550-3213(84)90052-X}{{\em Nucl.Phys.}
  {\bfseries B241} (1984) 333--380}.

\bibitem{Mattis:1986mj}
M.~P. Mattis, ``{Correlations in Two-dimensional Critical Theories},''
\href{http://dx.doi.org/10.1016/0550-3213(87)90361-0}{{\em Nucl.Phys.}
  {\bfseries B285} (1987) 671}.

\bibitem{Ginsparg:1988ui}
P.~H. Ginsparg, ``{Applied Conformal Field Theory},''
\href{http://arxiv.org/abs/hep-th/9108028}{{\ttfamily arXiv:hep-th/9108028
  [hep-th]}}.

\end{mcitethebibliography}\endgroup

\end{document}